\documentclass[prl,aps,twocolumn]{revtex4-1}
\usepackage{lmodern}
\usepackage{lmodern}

\usepackage[T1]{fontenc}
\usepackage[latin9]{inputenc}
\usepackage{babel}
\usepackage{amsmath}
\usepackage{amssymb}
\usepackage{graphicx}
\usepackage[unicode=true,pdfusetitle,
 bookmarks=true,bookmarksnumbered=false,bookmarksopen=false,
 breaklinks=false,pdfborder={0 0 1},backref=false,colorlinks=false]
 {hyperref}

\makeatletter
\usepackage{babel}

\makeatother

\begin{document}
\title{Spatial distribution of local elastic moduli in nanocrystalline metals}
\author{Menahem Krief}
\email{menahem.krief@mail.huji.ac.il}

\author{Yinon Ashkenazy}
\address{Racah Institute of Physics, The Hebrew University, 9190401 Jerusalem,
Israel}
\begin{abstract}
Elastoplastic properties of nanocrystalline metals are non-uniform on the scale of the grain size, and this non-uniformity affects macroscopic quantities  as, in these systems, a significant part of the material is at or adjacent to a grain boundary.
 We use molecular dynamics simulations to study the spatial distributions of local elastic moduli in nano-grained pure metals and analyze their dependence on grain size.
 Calculations are performed for
copper and tantalum with grain sizes ranging from 5-20nm. Shear modulus distributions for grain and grain-boundary atoms were calculated.
It is shown that the non-crystalline grain boundary has a wide shear-modulus
distribution, which is grain-size independent, while grains have a
peaked distribution, which becomes sharper with increasing grain size.
Average elastic moduli of the bulk, grains, and grain boundary are
calculated as a function of grain size. The atomistic simulations
show that the reduction of total elastic moduli with decreasing
grain size is mainly due to a resulting larger grain-boundary atoms
fraction, and that the total elastic moduli can be approximated by
a simple weighted average of larger grains elastic moduli and a
lower grain-boundary elastic moduli.
\end{abstract}
\maketitle

\section{Introduction}

Understanding the elastic properties of metals plays
a key role in the analysis and modeling of the response to deformation under varying conditions.
The local elastic properties in polycrystalline metals depend on the local structure, which varies from crystalline to grain boundaries
and hetero-phase interfaces \cite{gleiter2000nanostructured,schiotz2003maximum,wolf2005deformation,keblinski1999structure, antolovich2014plastic, mathaudhu2020building, zepeda2017probing,liu2023stresses}. 
The local variation of elastic properties 
plays a key role in tailoring of the macroscopic effective material properties, specifically in cases where grain refinement is used \cite{mathaudhu2020building}. Similarly local properties control strengthening in composite materials whether produced by additive manufacturing \cite{zhang2023additive} or via self organized segregation (e.g. \cite{kim2022strengthening}) as well as in developing models for structure evolution \cite{pant2022role}.
In order to develop an understanding of global response functions for composite materials, it is necessary to describe detailed distributions of local properties in addition to global averages.
Since local elastic properties are not accessible experimentally for a wide range of systems, it is beneficial to study surrogate model systems, which allow reliable numerical evaluation of these.  
Such simulations play a key role in developing effective models for composite materials. 

In Ref. \cite{krief2021calculation} we demonstrated the feasibility,
robustness, and accuracy of calculating bulk elastic constants
using molecular dynamics simulations in the NVT ensemble (where the number of particles, volume and temperature are kept constant). 
This method
is generalized in this work to include the calculation of local elastic properties.
The main advantage of this approach is that all components of the
elasticity tensor are obtained in a single consistent molecular-dynamics
simulation, as opposed to the standard explicit deformation method,
which requires several simulations under different deformations and
is limited to evaluating global average values. Using NVT local evaluators,
calculations are inherently local and immediate, so localized averaging leading to spatial and temporal distributions is trivial. In addition, while special care needs to be taken in the direct drive method to avoid strain rate and defect formation effects, the NVT local evaluation does not rely on forcing deformation. It thus is not affected by these potential problems. 
Moreover, in the standard deformation method, a non-negligible deformation of the simulation box is required  in order to create a measurable variation in local stress. In inhomogeneous systems where elastic properties vary locally, such deformations lead to strain localization, which in many cases results in local yield \cite{shimizu2007theory,zepeda2017probing} that prevents the calculation of elastic moduli.

Previously, direct calculations
of local elastic constants by atomistic simulations were performed
in Refs. \cite{kluge1990formalism,lutsko1989generalized} for copper
and gold, in Refs \cite{yoshimoto2004mechanical, tsamados2009local} for amorphous polymeric
glasses, in Ref. \cite{lips2018stress} for a lipid bilayer and in
Ref. \cite{veldhorst2018mechanical} for ionic liquids. In Ref. \cite{magazzeni2021nanoindentation}
an experimental correlative approach was used to extract the local elastic
properties of titanium. 

In this paper, we present and analyze the distribution of
local elastic constants in nanocrystalline copper and tantalum, as calculated using
molecular dynamics simulations employing realistic many-body potentials.
We extract the distributions of elastic moduli within polycrystalline systems
and study their dependence on  grain size in the range
of 5-20 nm. In addition, we study the grain-size dependence of the
average elastic moduli of the total system, in grains and grain boundary,
and show explicitly that average global values agree with mean field models that were used in the literature
\cite{shen1995elastic,schiotz1998softening,latapie2003effect,pan2008tensile,chen2009size,valat2017grain}.
We note that existing models for the calculation of grain-boundary elastic properties of nanoscrystalline metals, usually rely on such effective mean field models, which assume a grain size independent grain-boundary moduli \cite{gao2013studying}.
  Such methods only allow the calculation of global averages. The method which we employ in this work, gives a direct evaluation of local elastic properties, without relying on some of the major assumptions which were previously made - i.e. the description of the elastic properties of the grain boundary atoms using a single average value, or the independence of local properties on interface curvature. The method enables the calculation of new observables such as the distributions of elastic properties, which can play a crucial role in developing stochastic models addressing the response under non-deterministic drive conditions \cite{engelberg2018stochastic,pant2022role}.

\section{Local Elasticity\label{sec:Thermoelasticity}}

In this section, we outline the method for the calculation
of local elastic constants using molecular dynamics simulations in the NVT ensemble.

It was shown in previous works \cite{squire1969isothermal,lutsko1989generalized,ccaugin1999thermal,yoshimoto2004mechanical,van2005isothermal,barrat2006microscopic,clavier2017computation,lips2018stress,krief2021calculation,tadmor2011modeling},
that in the NVT ensemble, the elasticity tensor can be written as
a sum of three contributions: (i) a purely configurational part known
as the Born term, which is given by a canonical average of the second
order derivative of the potential energy with respect to the Lagrangian
strain tensor, (ii) a stress fluctuation term that vanishes at zero
temperature and (iii) a kinetic ideal gas contribution, which also
vanishes at zero temperature. As a result, the low-temperature limit
for the total elasticity tensor can be written on a per-atom basis
\cite{alber1992grain,tadmor2011modeling}, in the form:

\begin{align}
C_{\alpha\beta\gamma\delta} & =\left\langle \frac{1}{V}\sum_{i}V_{i}C_{i,\alpha\beta\gamma\delta}^{B}\right\rangle ,\label{eq:elastic_micro}
\end{align}
where $\alpha,\beta,\gamma,\delta$ are the directional tensor indices,
$\left\langle \cdot\right\rangle $ represents ensemble average, $C_{i,\alpha\beta\gamma\delta}^{B}$
is the local, per-atom Born elasticity tensor, $V$ is the total system
volume and $V_{i}$ is the volume of atom $i$, defined such that
$V=\sum_{i}V_{i}$. Following Refs. \cite{alber1992grain,tadmor2011modeling}, we
define the local volume of a specific atom as the volume of the Voronoi cell associated
with it. 
In this work, calculations were performed for copper and
tantalum, modeled by embedded-atom-model (EAM) potentials \cite{daw1984embedded}.
These potentials are defined by a pair potential function
$v=v\left(r\right)$, an embedding function $F=F\left(\rho\right)$
and a local density function $\rho=\rho\left(r\right)$, so that the the potential energy takes the form:
\begin{equation}
E\left(\boldsymbol{r}_{1},...,\boldsymbol{r}_{N}\right)=\sum_{i}F\left(\rho_{i}\right)+\sum_{i<j}v\left(r_{ij}\right),\label{eq:EAM}
\end{equation}
where $\boldsymbol{r}_{i}$ is the position of atom $i$, $\boldsymbol{r}_{ij}=\boldsymbol{r}_{i}-\boldsymbol{r}_{j}$
and $\rho_{i}=\sum_{j\neq i}\rho\left(r_{ij}\right)$ is the density
around atom $i$. It can be shown that for an EAM potential, the per-atom
Born elasticity tensor takes the form \cite{lutsko1989generalized,wolf1992temperature,chantasiriwan1996higher,ccaugin1999thermal,tadmor2011modeling,krief2021calculation}:
\begin{align}
V_{i}C_{i,\alpha\beta\gamma\delta}^{B} & =\frac{1}{2}\sum_{j\neq i}X_{ij}\frac{r_{ij,\alpha}r_{ij,\beta}r_{ij,\gamma}r_{ij,\delta}}{r_{ij}^{2}}\nonumber \\
 & +F''\left(\rho_{i}\right)g_{i,\alpha\beta}g_{i,\gamma\delta}\label{eq:cborn_micro}
\end{align}
where:

\begin{align}
X_{ij} & =v''\left(r_{ij}\right)-\frac{1}{r_{ij}}v'\left(r_{ij}\right)\nonumber \\
 & +\left(F'\left(\rho_{i}\right)+F'\left(\rho_{j}\right)\right)\left(\rho''\left(r_{ij}\right)-\frac{1}{r_{ij}}\rho'\left(r_{ij}\right)\right),\label{eq:Xij}
\end{align}
and:
\begin{equation}
g_{i,\alpha\beta}=\sum_{j\neq i}\rho'\left(r_{ij}\right)\frac{r_{ij,\alpha}r_{ij,\beta}}{r_{ij}}.\label{eq:g_def}
\end{equation}
Various local elastic moduli can be obtained directly from the local
elasticity tensor \eqref{eq:cborn_micro}, by employing local Voigt
averages \cite{hill1952elastic,nye1985physical,newnham2005properties},
which results in the following expressions for the local Bulk modulus:

\begin{equation}
9B=C_{11}+C_{22}+C_{33}+2\left(C_{12}+C_{23}+C_{31}\right),\label{eq:B}
\end{equation}
the local Shear modulus:
\begin{align}
15G & =C_{11}+C_{22}+C_{33}-\left(C_{12}+C_{23}+C_{31}\right)\nonumber \\
 & +3\left(C_{44}+C_{55}+C_{66}\right),\label{eq:G}
\end{align}
the local Young's modulus:

\begin{equation}
\frac{1}{E}=\frac{1}{3G}+\frac{1}{9B},
\end{equation}
and the local Poisson's ratio:

\begin{equation}
\nu=\frac{1}{2}\left(1-\frac{3G}{3B+G}\right).\label{eq:poiss}
\end{equation}
The Voigt notation for tensor indices was used in equations \eqref{eq:B}-\eqref{eq:G}.

\section{Results\label{sec:Results}}

\begin{figure}
\begin{centering}
\includegraphics[scale=0.5]{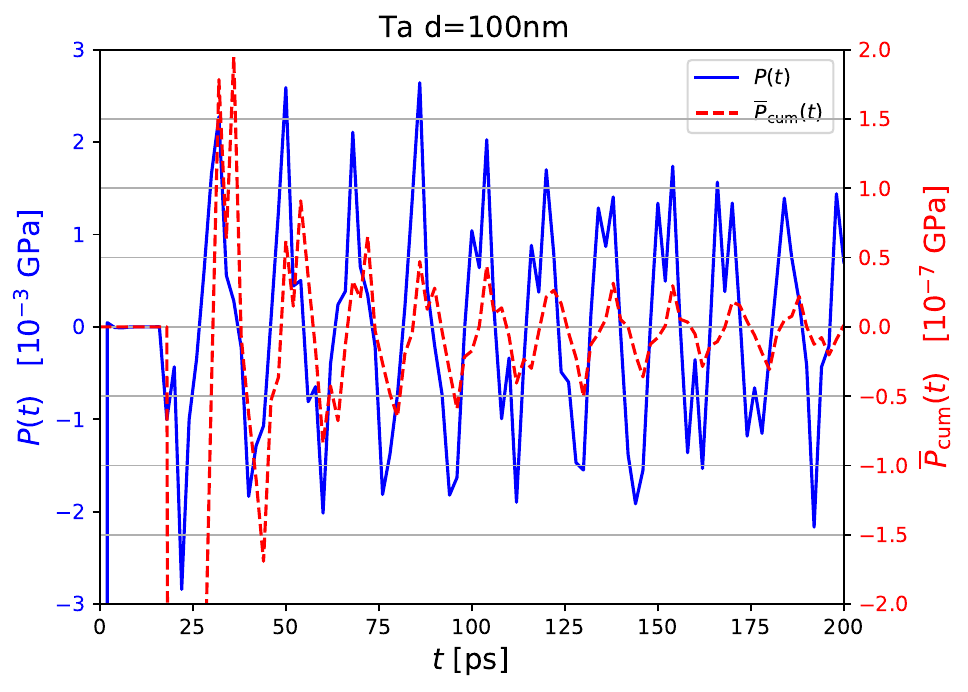} 
\par\end{centering}
\caption{Demonstration of the equilibration process for nanocrystalline tantalum
with grain size $d=100$nm. The box size is relaxed in the NPT ensemble,
until the total pressure is zero. Shown are the total instantaneous
pressure (blue solid line, left y-axis) and the average pressure (red
dashed line, right y-axis), which is calculated over a window of $10^{4}$
times steps (with a time step of $2$fs).
\label{fig:p_t_ta}}
\end{figure}

\begin{figure*}
\begin{centering}
\includegraphics[scale=0.2]{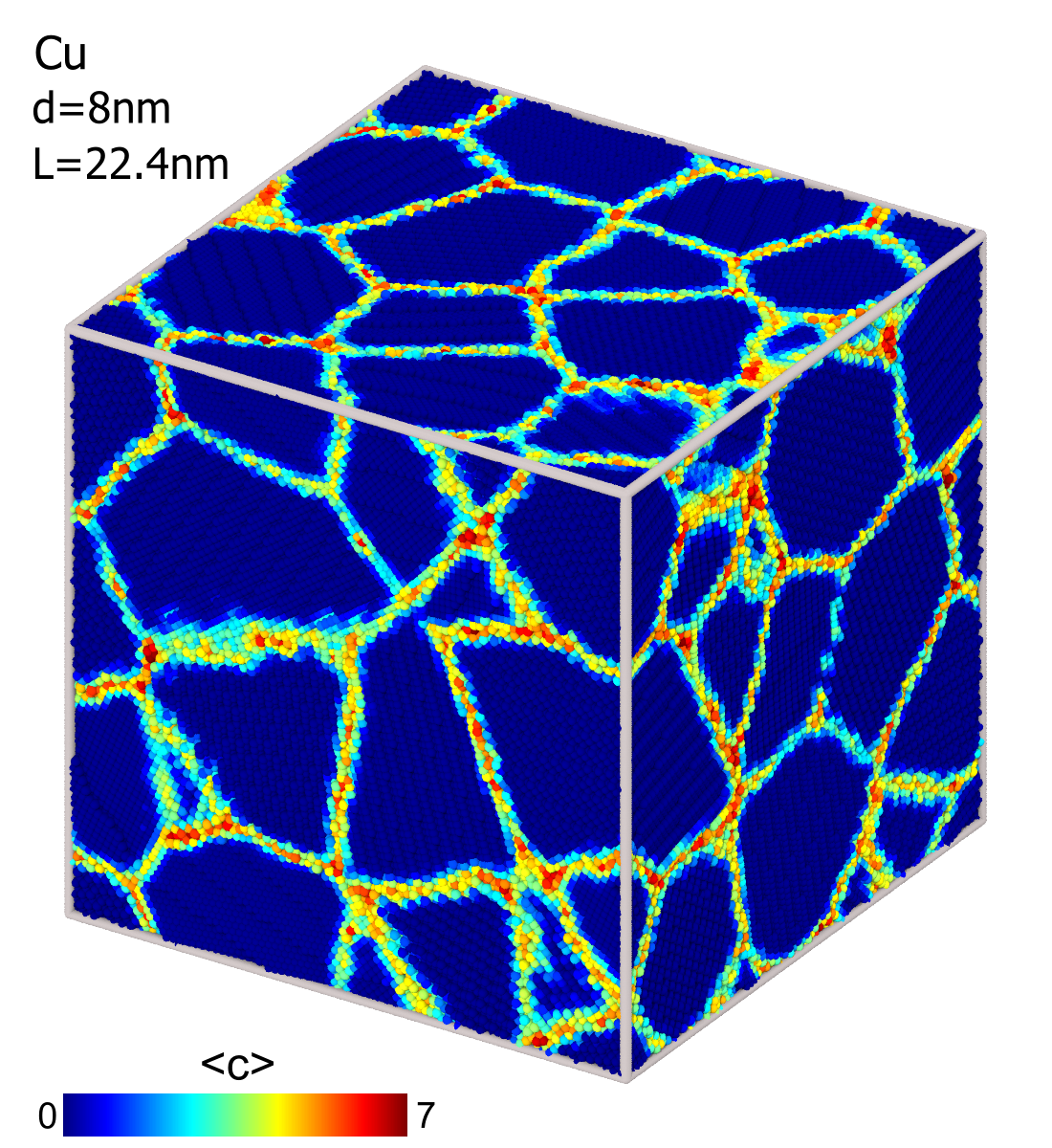}\includegraphics[scale=0.2]{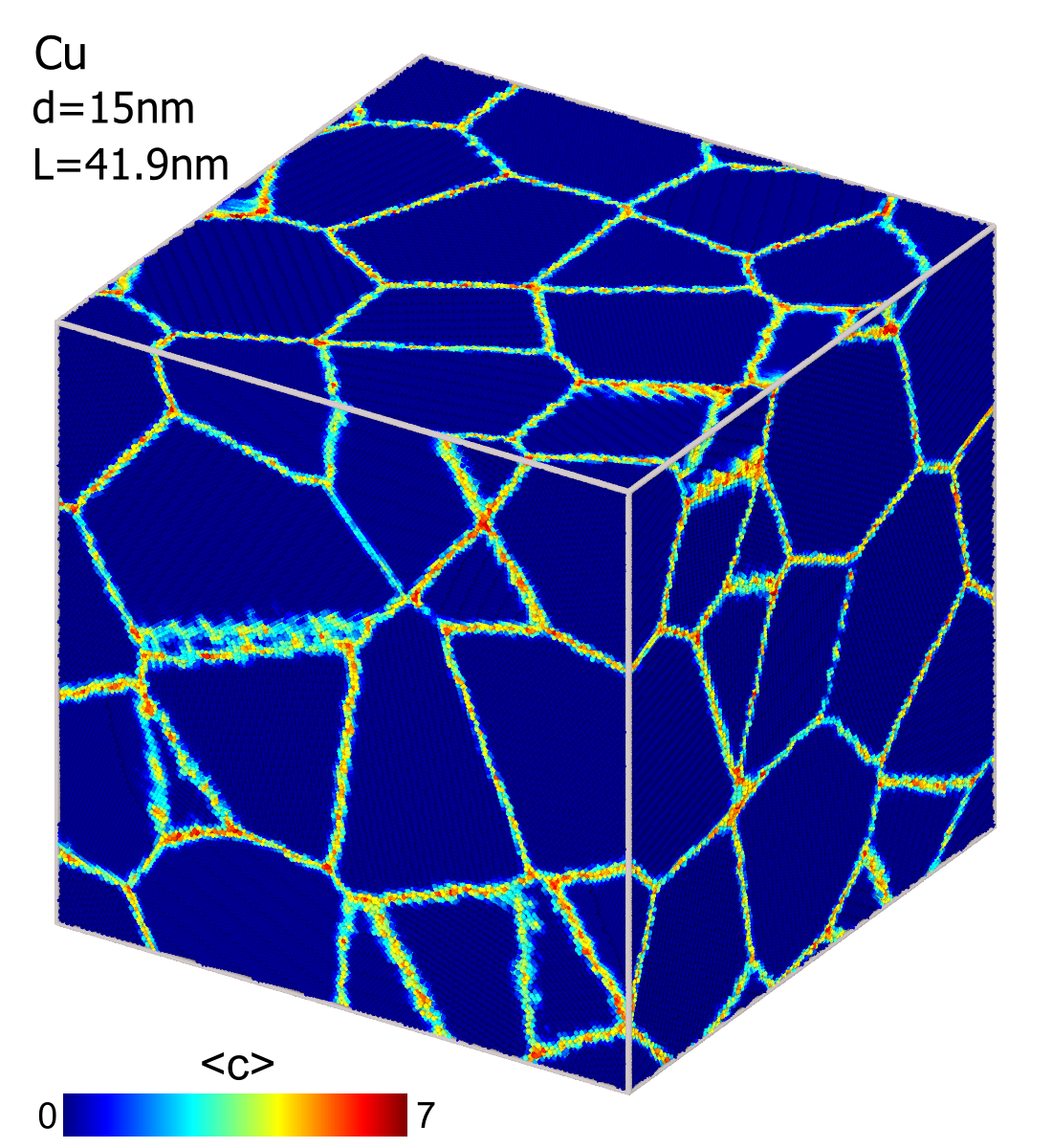}\includegraphics[scale=0.2]{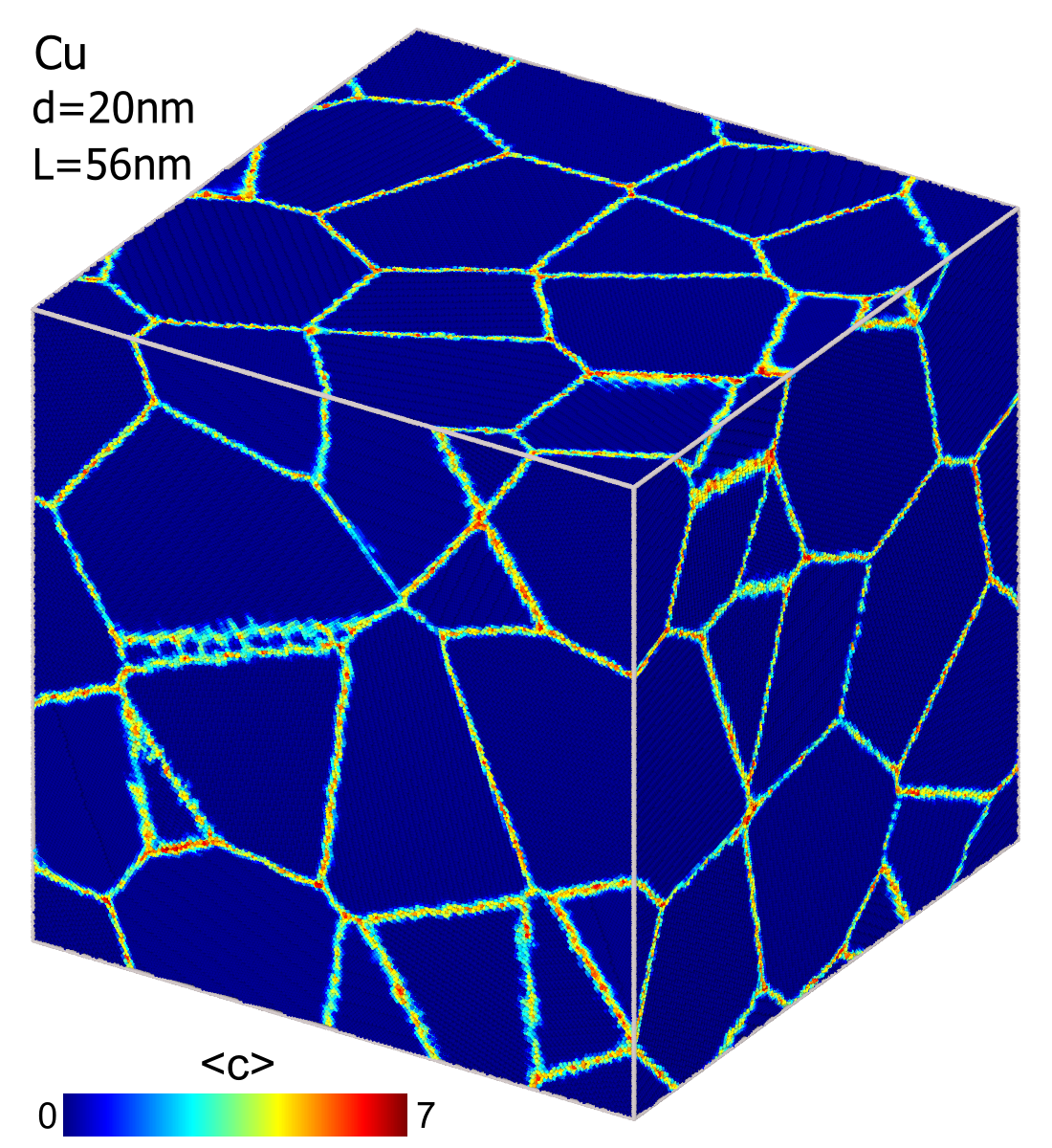} 
\par\end{centering}
\begin{centering}
\includegraphics[scale=0.2]{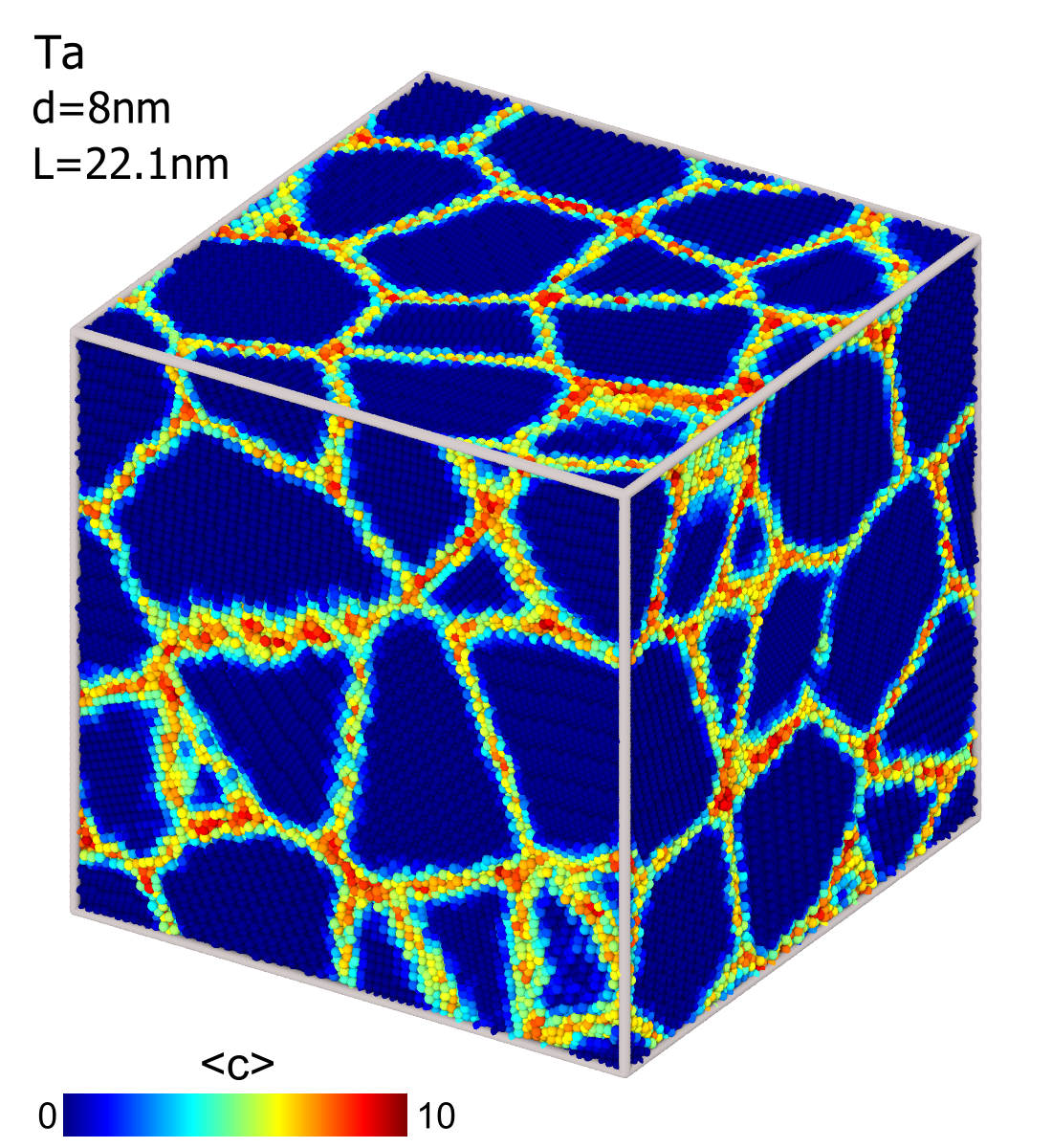}\includegraphics[scale=0.2]{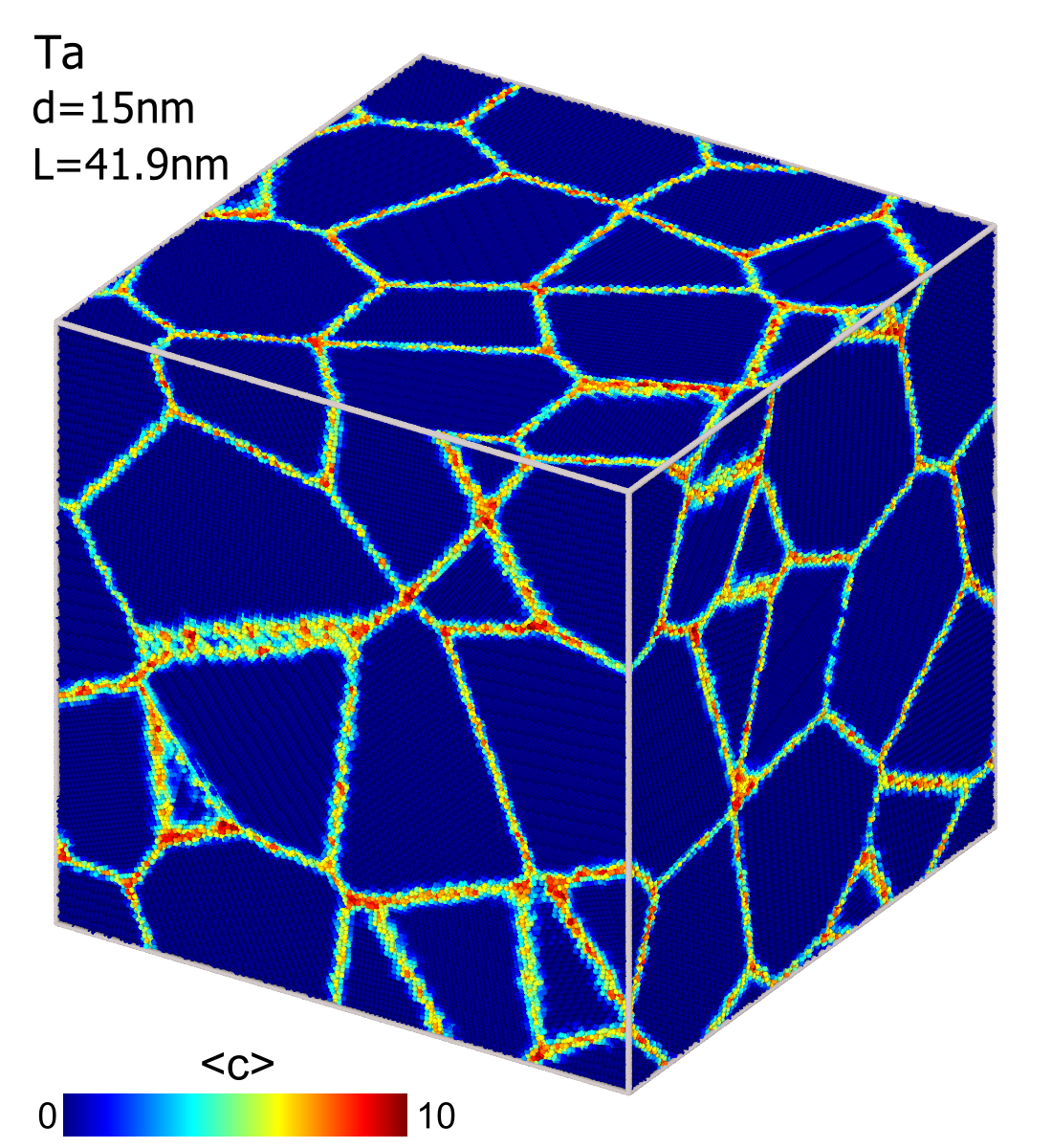}\includegraphics[scale=0.19]{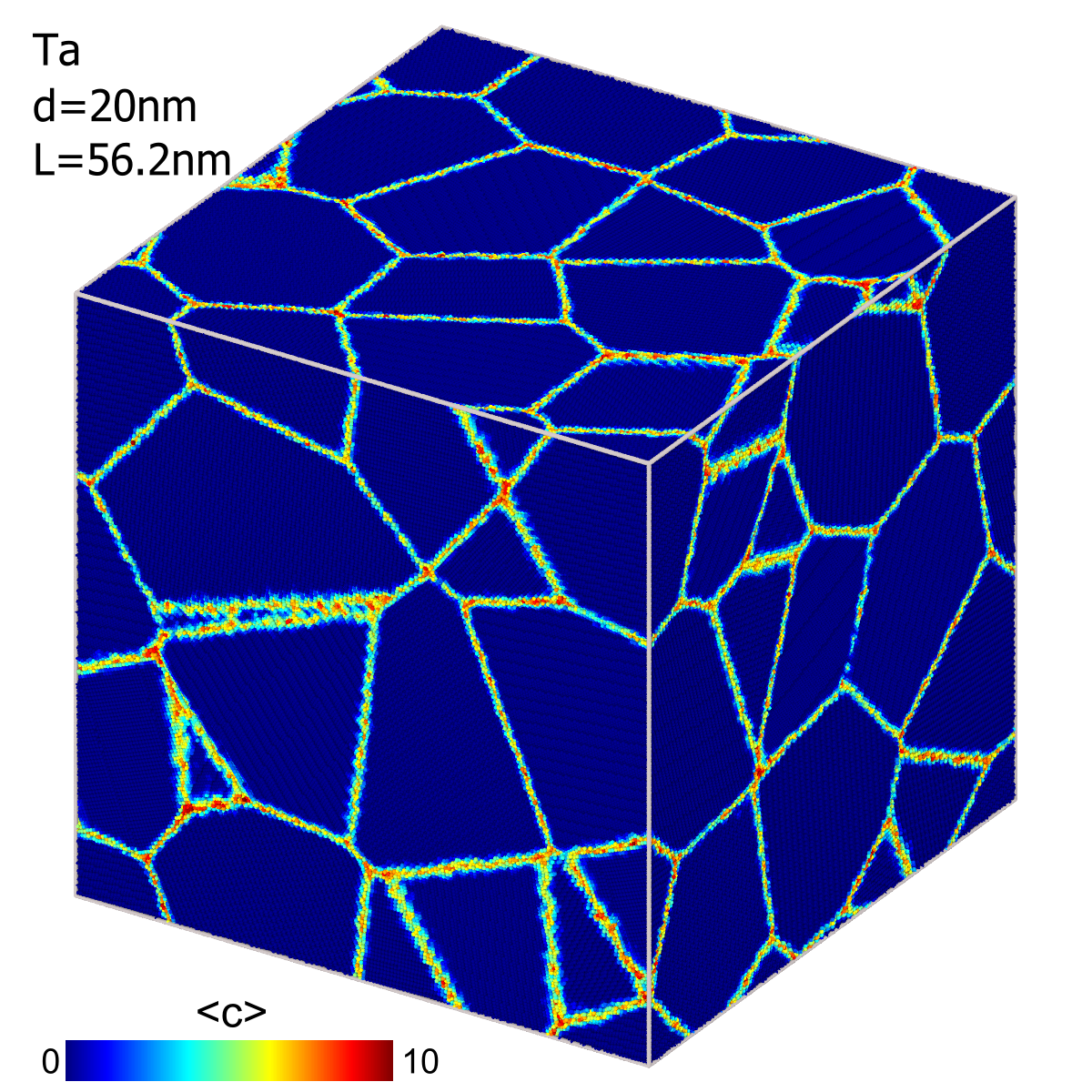} 
\par\end{centering}
\caption{Atomistic visualizations of simulated nanocrystalline copper (FCC,
upper panes) and tantalum (BCC, lower panes), with grain sizes $d=8$nm
(left panes), $15$nm (middle panes) and $20$nm (right panes). The
corresponding box size $L$ is also listed in each pane. Color indicates
the local average centrosymmetry parameter. It is evident that the number of grains is kept constant for different grains sizes.\label{fig:boxes}}
\end{figure*}

\begin{figure*}
\begin{centering}
\includegraphics[scale=0.2]{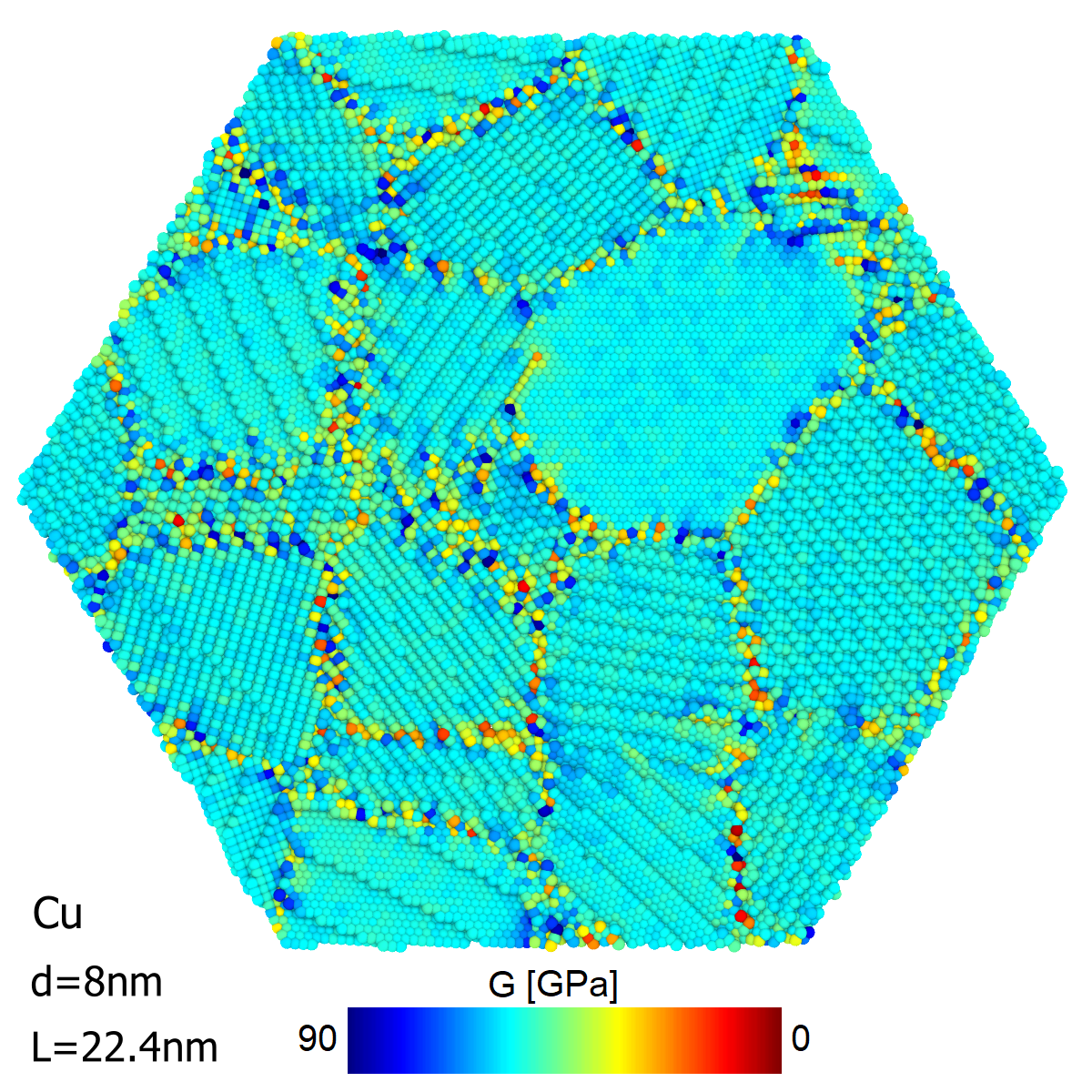}\includegraphics[scale=0.2]{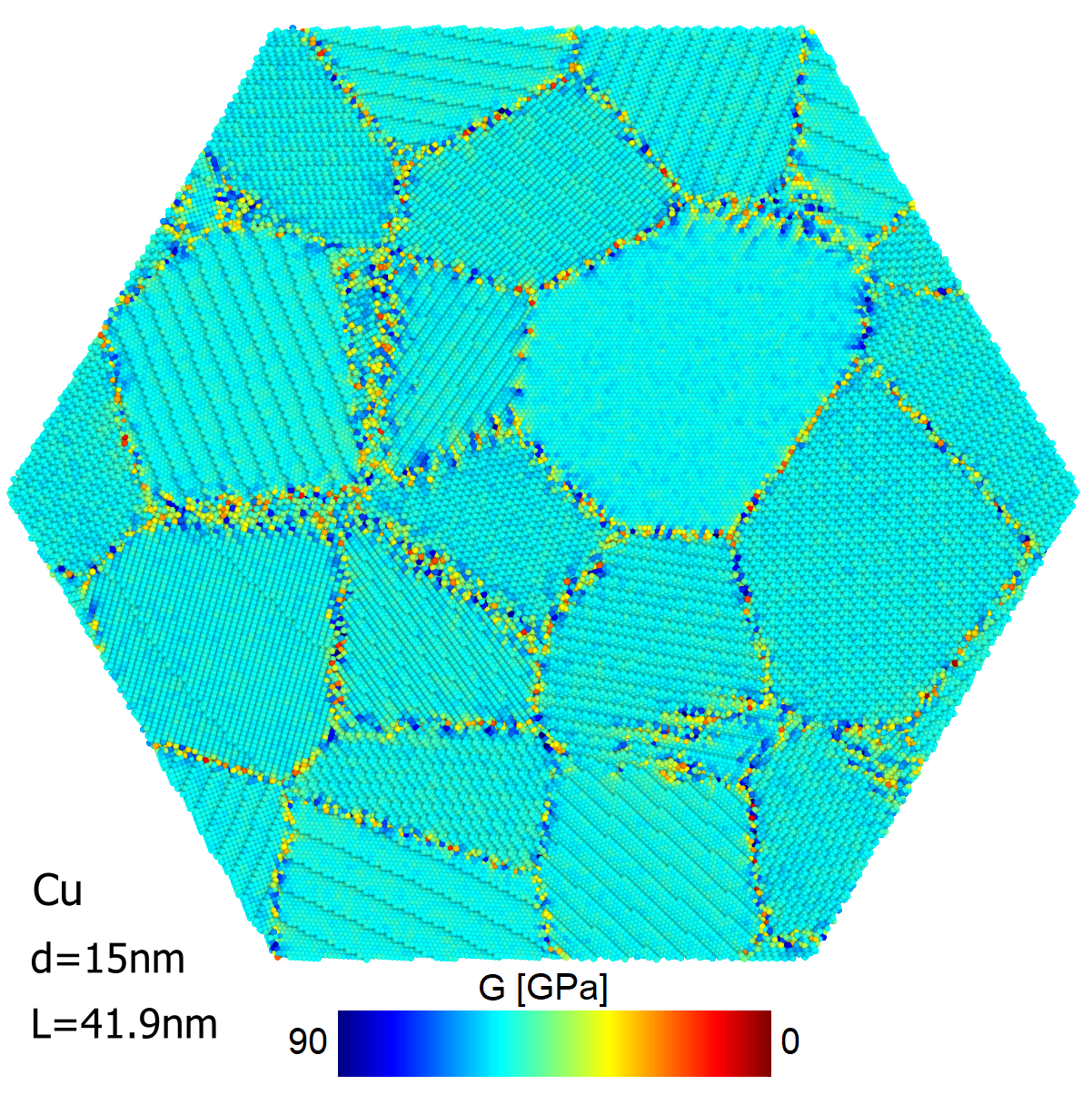}\includegraphics[scale=0.2]{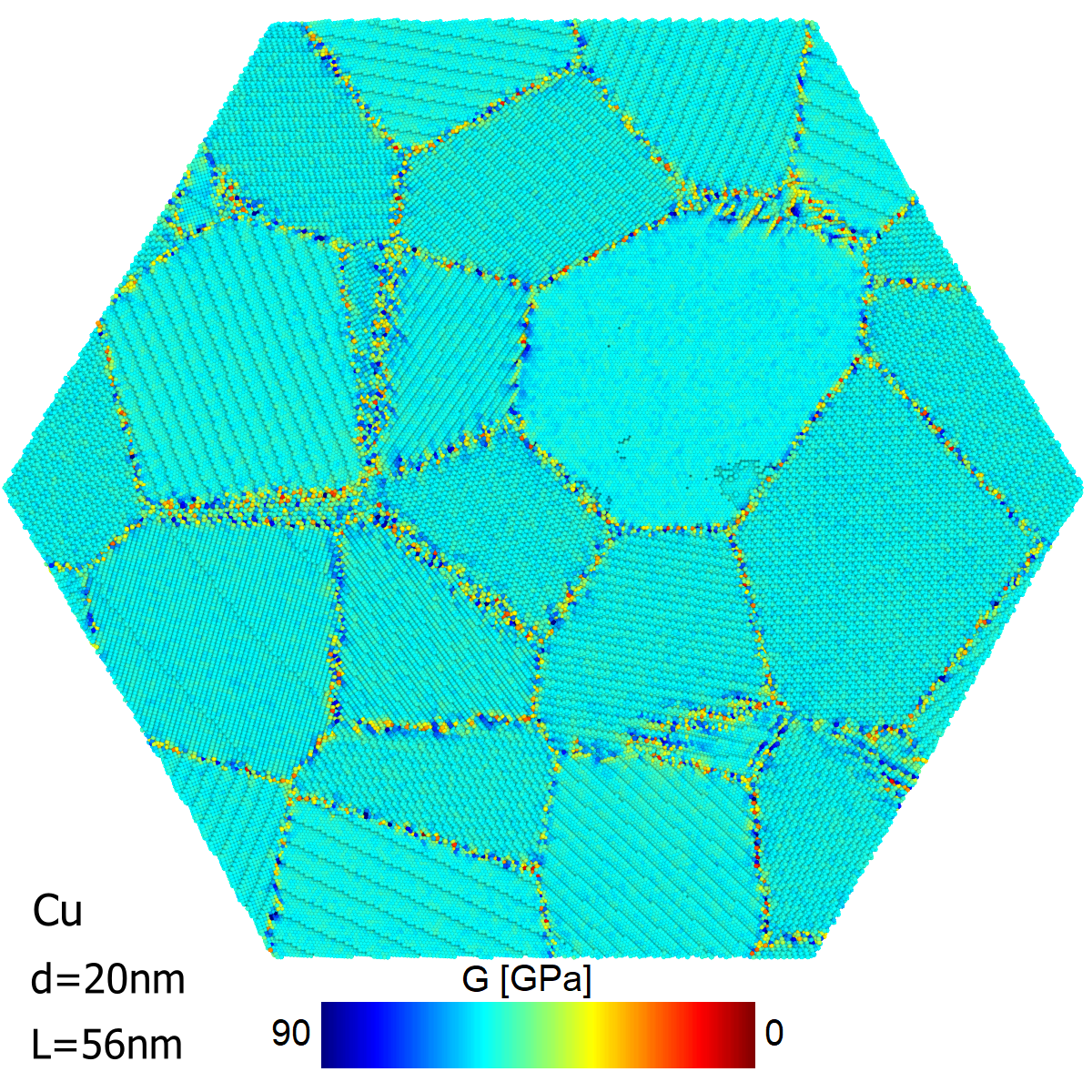} 
\par\end{centering}
\begin{centering}
\includegraphics[scale=0.2]{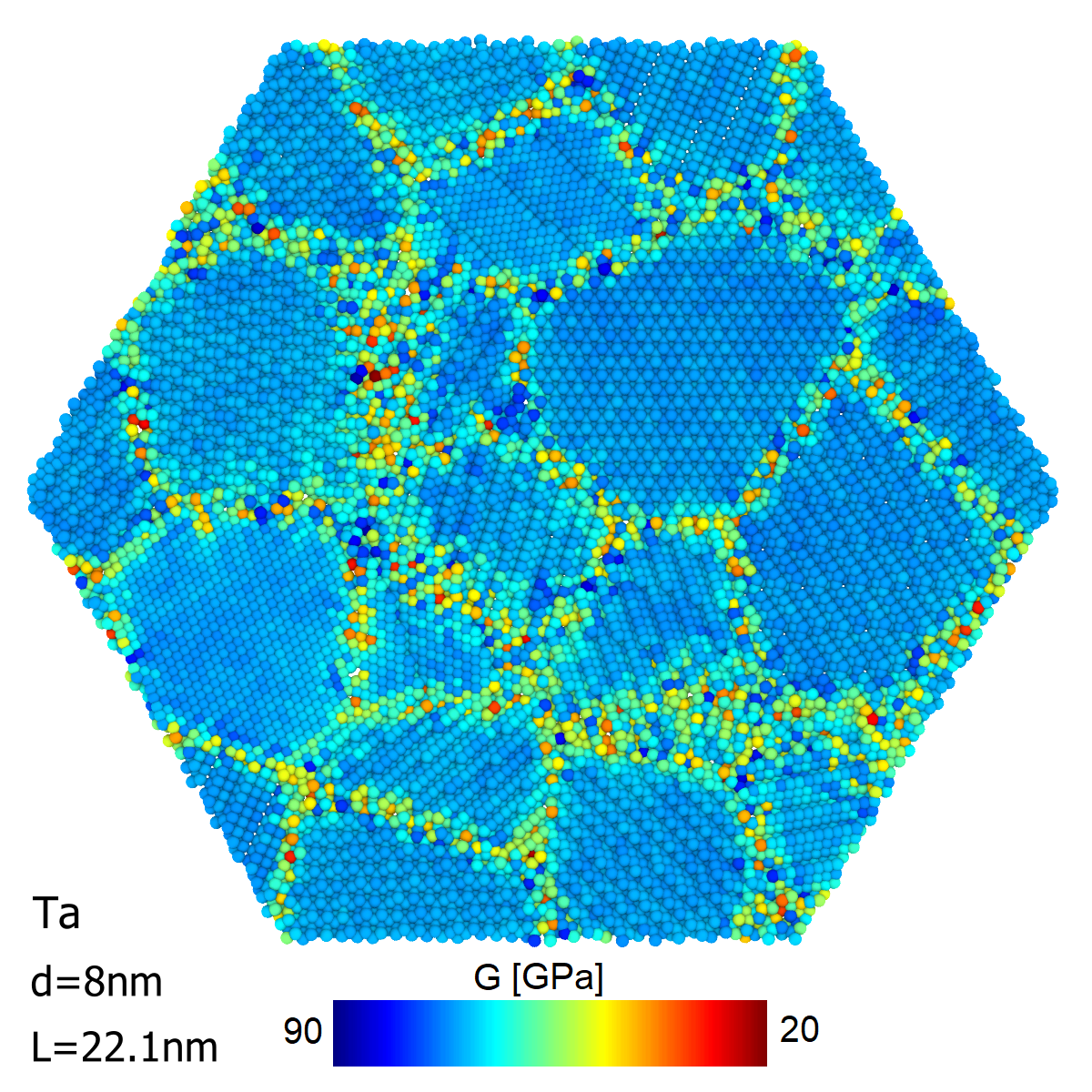}\includegraphics[scale=0.2]{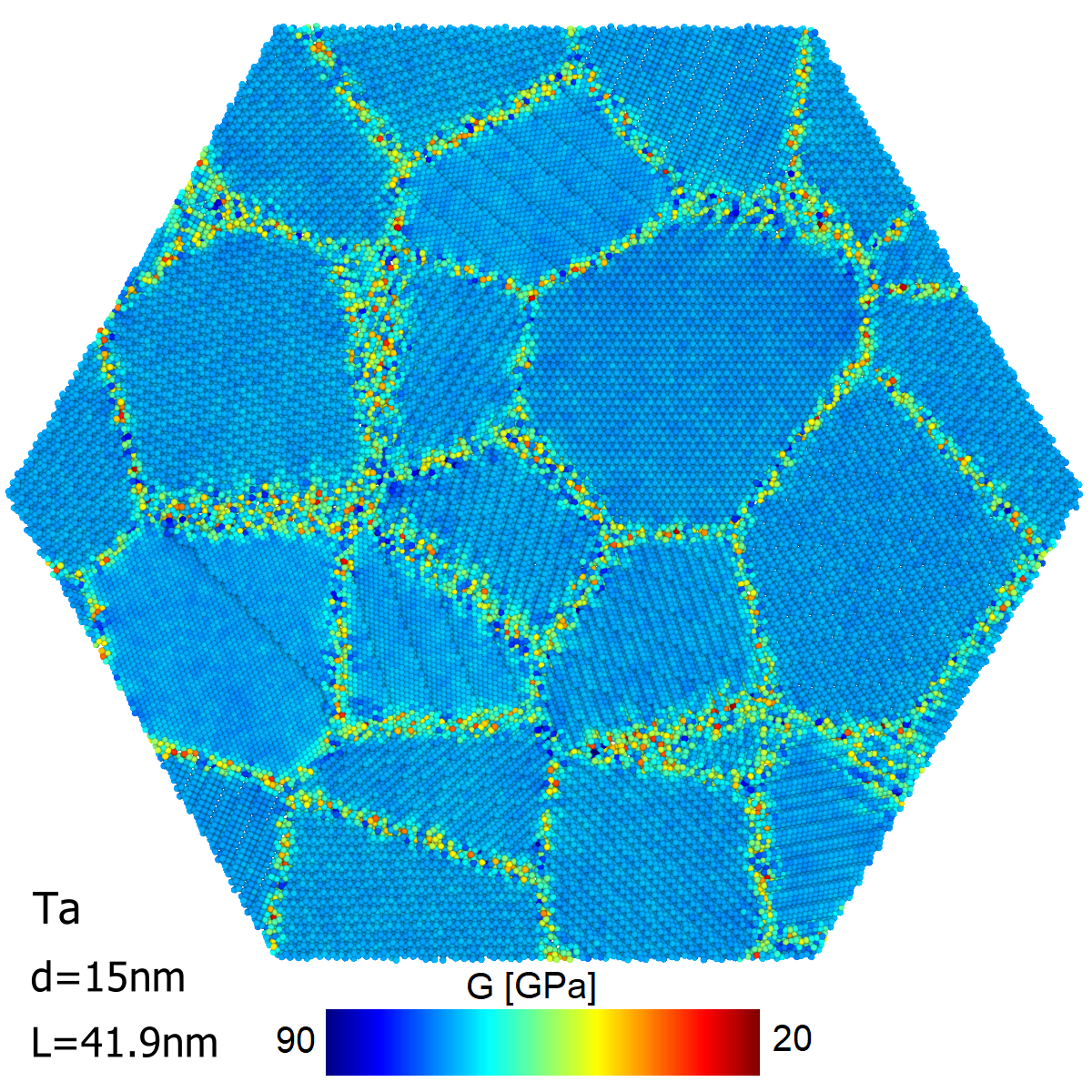}\includegraphics[scale=0.2]{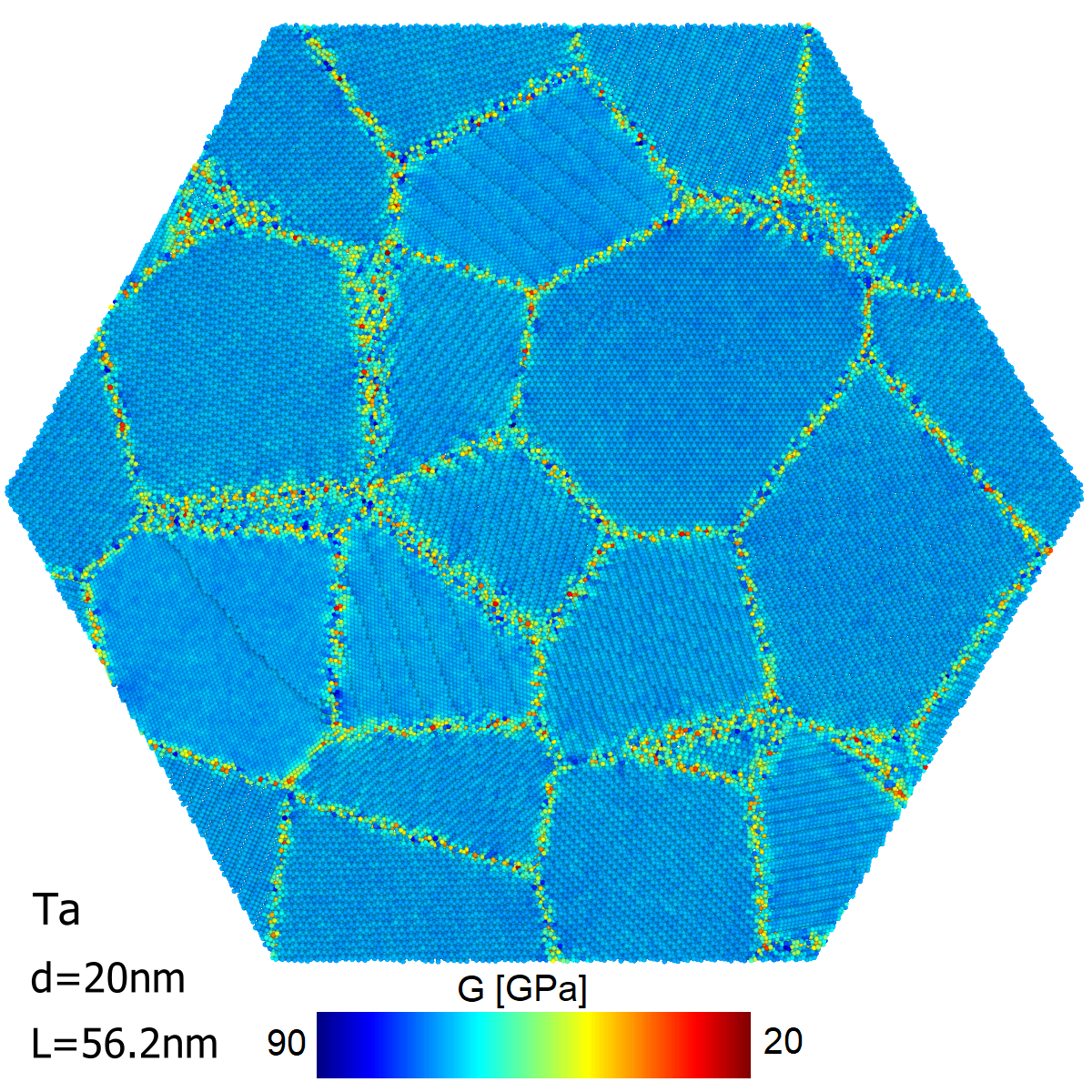} 
\par\end{centering}
\caption{Atomistic view of (111) planar slices of simulated nanocrystalline
copper (upper panes) and tantalum (lower panes), with grain sizes
$d=8$nm (left panes), $15$nm (middle panes) and $20$nm (right panes).
The corresponding simulation box size $L$ is also listed in each
pane. 
The spatial distribution of the local shear modulus $G$, is indicated by color. The lack of spatial correlation between the widely distributed values within the grain boundary atoms is evident. In contrast, deviations from the average in the grain are significantly smaller, as expected. It is also seen that the grain atoms have a larger value on average. These results are consistent with Fig. \ref{fig:histograms}. We note that the parallel lines in some of the grains are due to imgae shading, and are artificial.
\label{fig:snapshots}}
\end{figure*}

\begin{figure}
\begin{centering}
\includegraphics[scale=0.5]{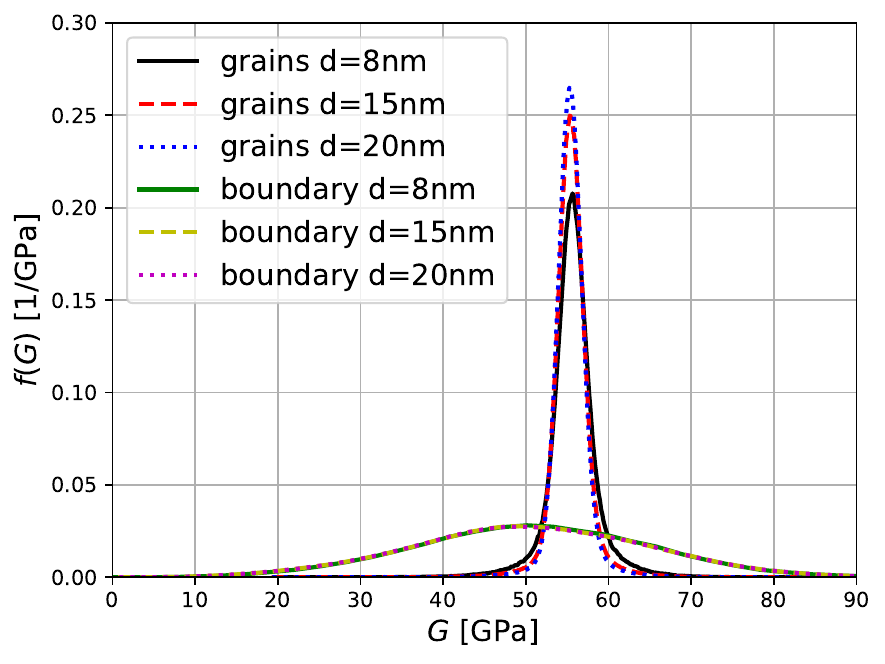} 
\par\end{centering}
\begin{centering}
\includegraphics[scale=0.5]{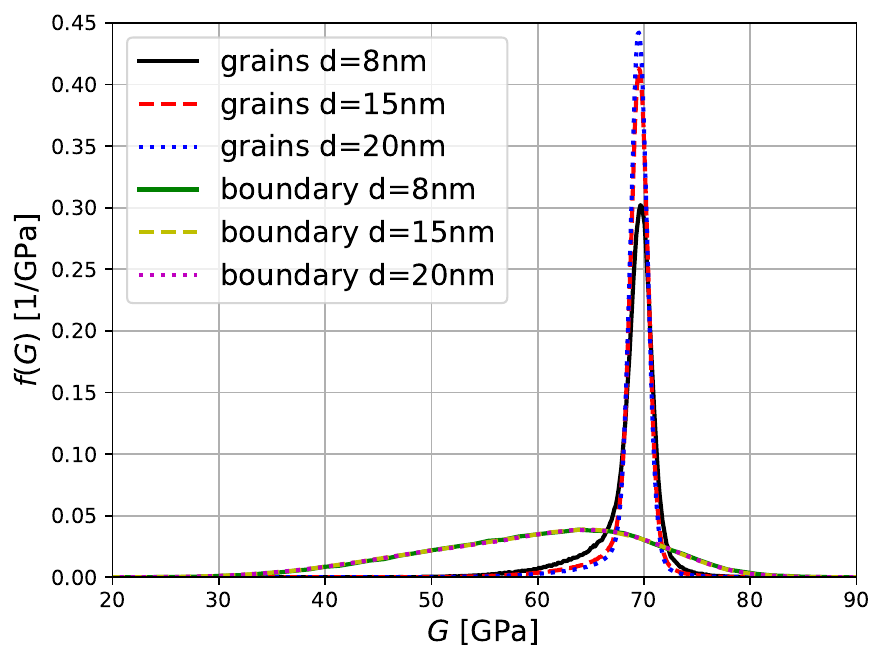} 
\par\end{centering}
\caption{Shear modulus probability densities for grain and grain-boundary atoms,
in nanocrystalline copper (upper pane) and tantalum (lower pane) and
for various grain sizes $d$ (as listed in the legend).
\label{fig:histograms}}
\end{figure}

\begin{figure}[t]
\begin{centering}
\includegraphics[scale=0.5]{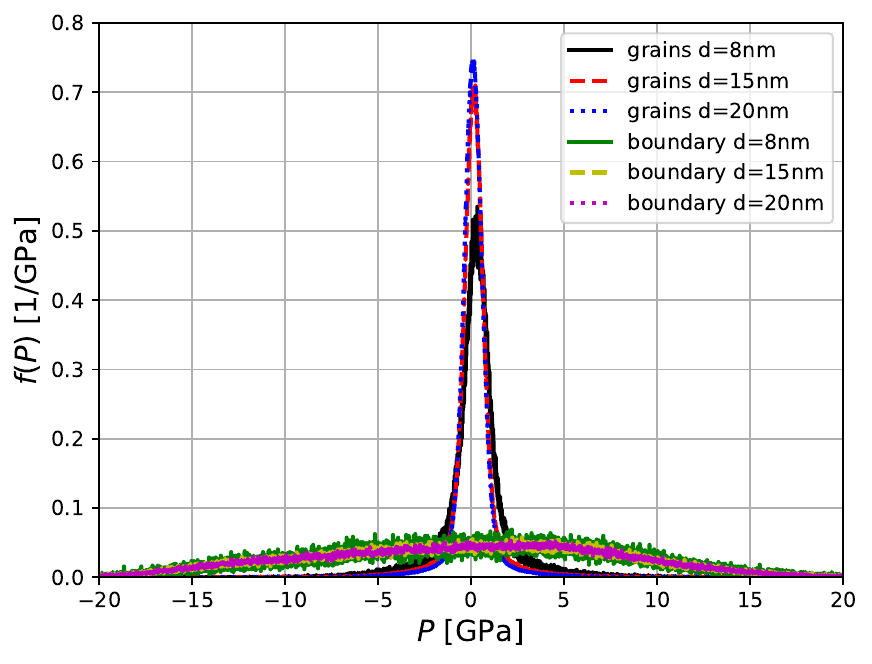}
\par\end{centering}
\caption{Grain and grain-boundary local pressure distributions in relaxed nanocrystalline
tantalum for various grain sizes $d$ (as listed in the legend).\label{fig:p_ta_histograms}}
\end{figure}

\begin{figure}
\begin{centering}
\includegraphics[scale=0.5]{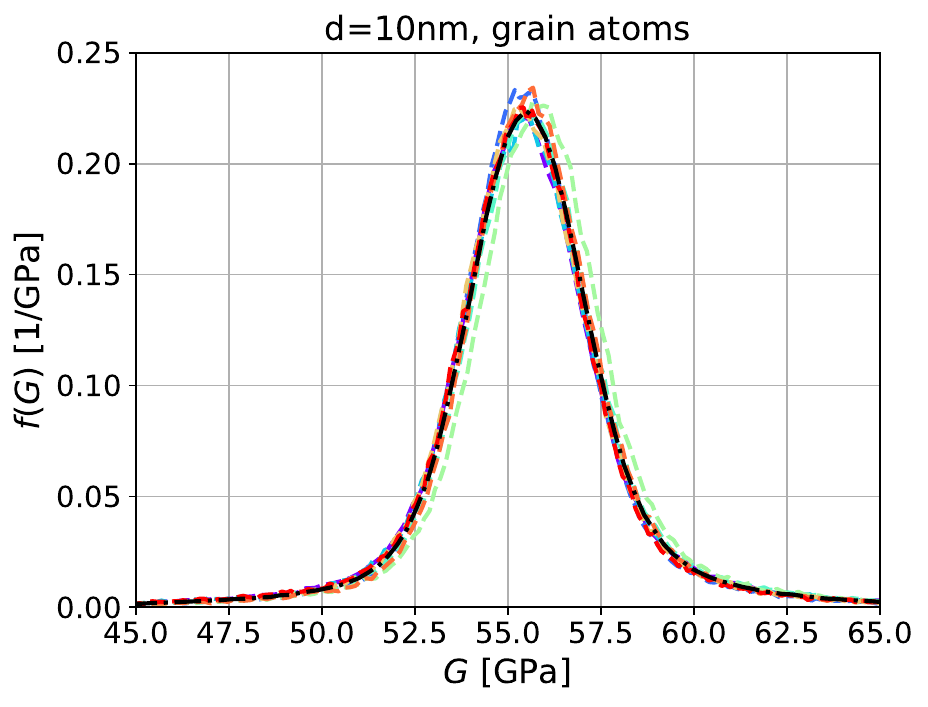} 
\par\end{centering}
\begin{centering}
\includegraphics[scale=0.5]{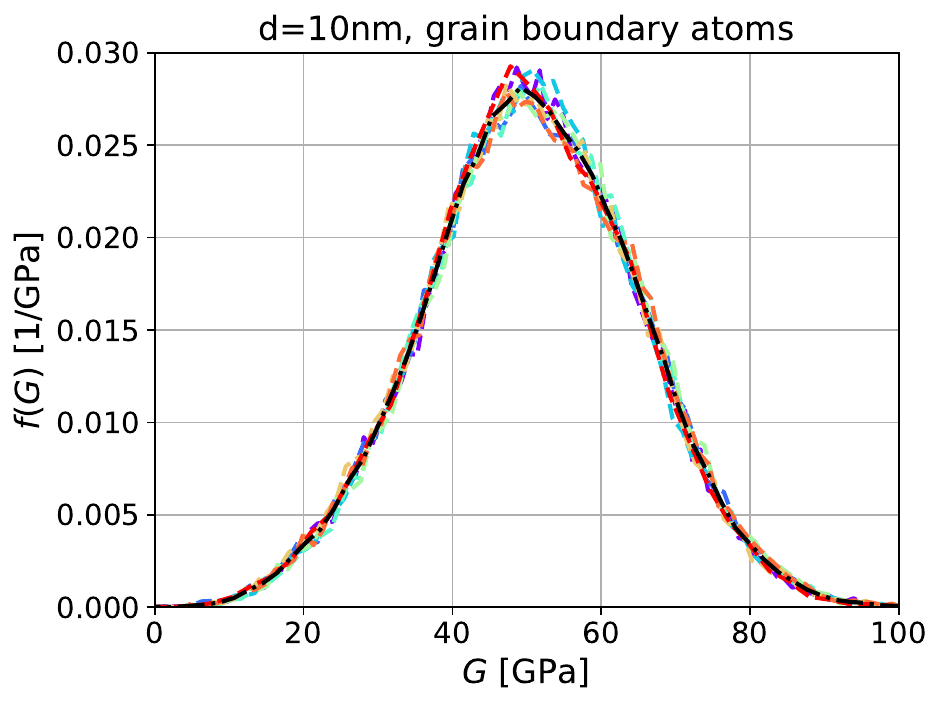} 
\par\end{centering}
\caption{A comparison of the shear modulus distributions in 8 different octants of the simulation box (colorful dashed lines) and in the entire system (black dotted lines). The results are shown for grain (upper pane) and grain-boundary (lower pane) atoms of nanocrystalline copper with grain size $d=10$nm. 
\label{fig:histograms_err}}
\end{figure}



\begin{figure}
\begin{centering}
\includegraphics[scale=0.5]{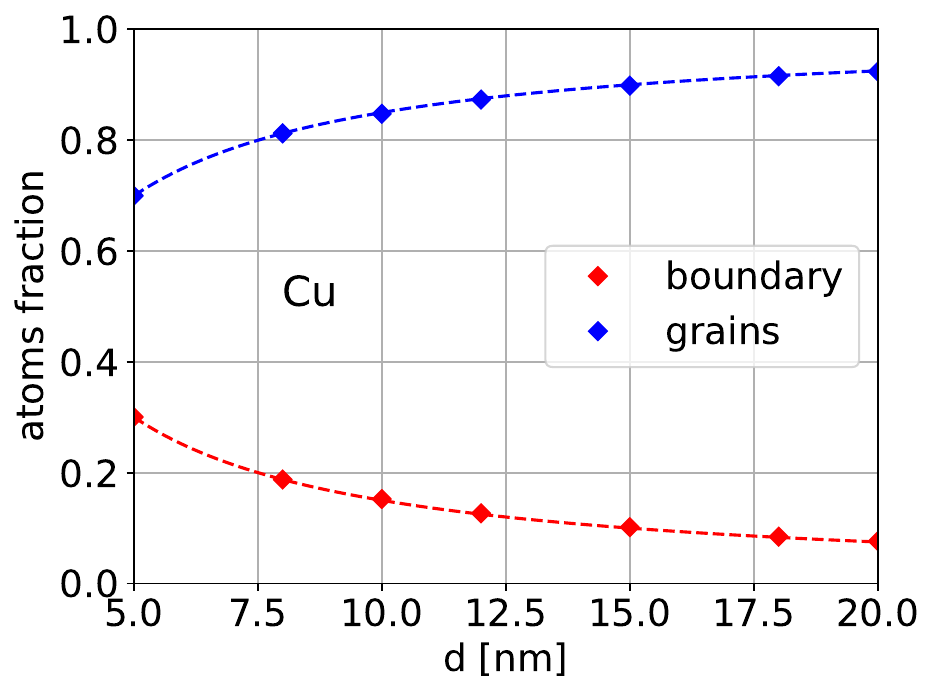} 
\par\end{centering}
\begin{centering}
\includegraphics[scale=0.5]{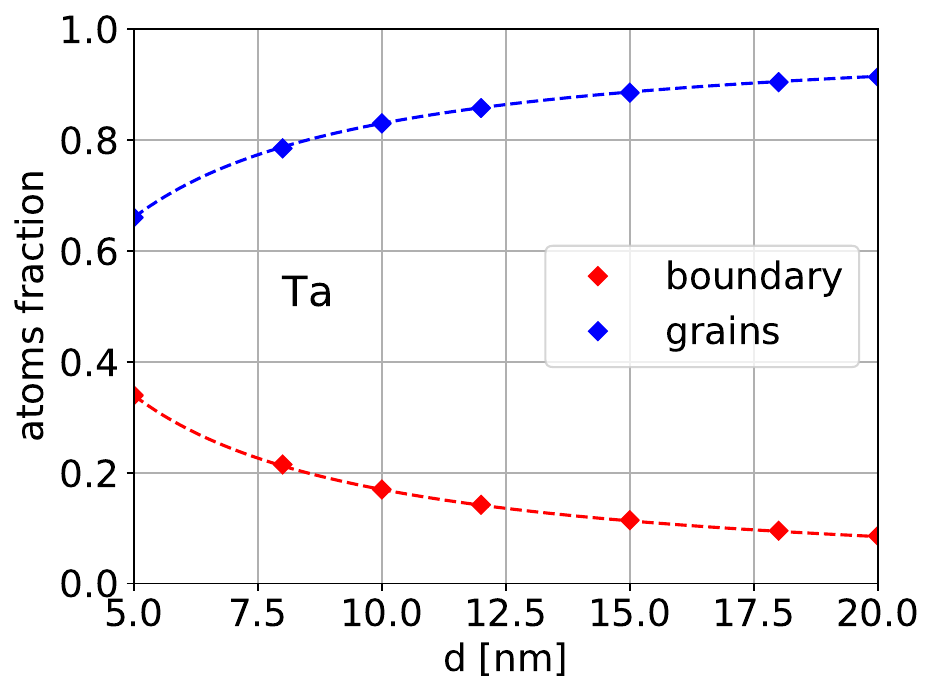} 
\par\end{centering}
\caption{
The fractional number of grain (in blue) and grain-boundary (in red) atoms as a function of grain size
in nanocrystalline copper (upper pane) and tantalum (lower pane). The points are the atomic fractions in the simulations (see Figure \ref{fig:boxes}), and the dashed lines are obtained from a $1/d$ fit (according to the scaling relation \eqref{eq:n_d_inv}, as discussed in the text).\label{fig:atfrac}}
\end{figure}

\begin{figure*}
\begin{centering}
\includegraphics[scale=0.28]{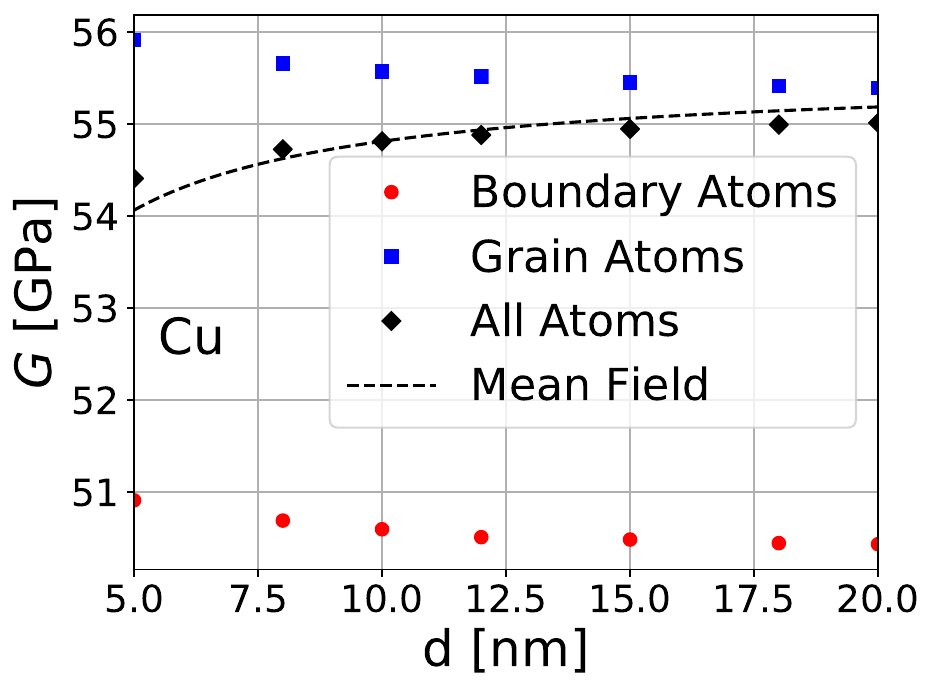}\includegraphics[scale=0.28]{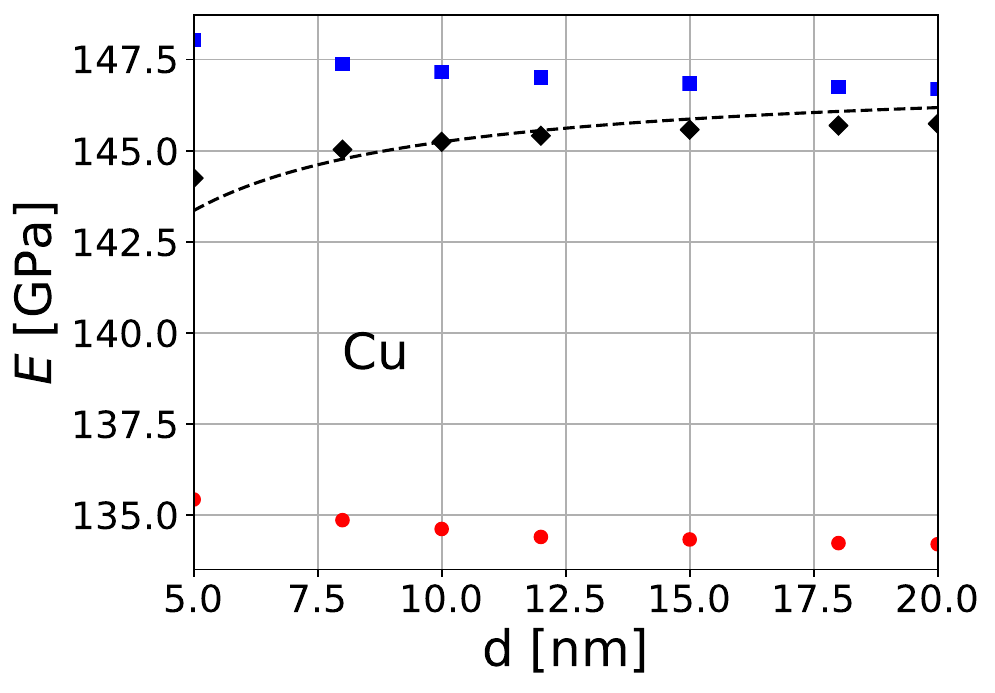}\includegraphics[scale=0.28]{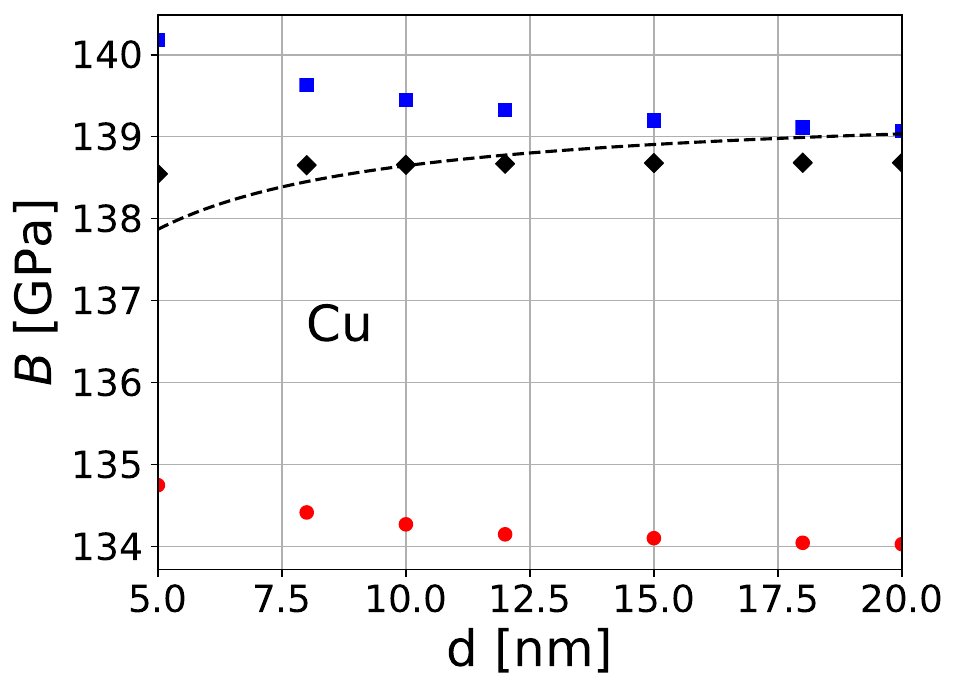}\includegraphics[scale=0.28]{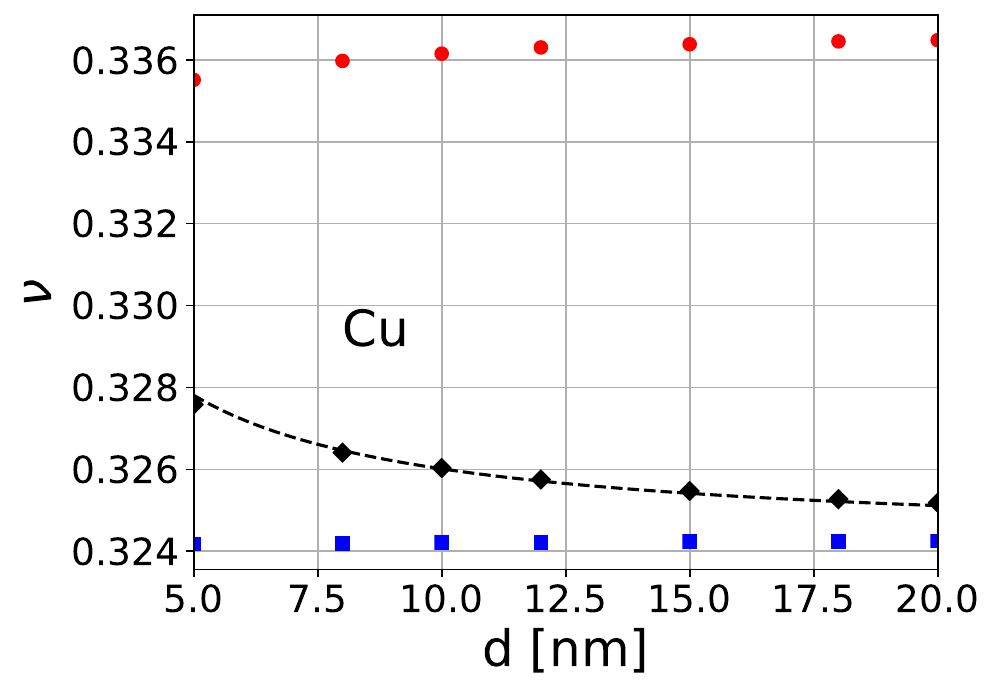}
\par\end{centering}
\begin{centering}
\includegraphics[scale=0.28]{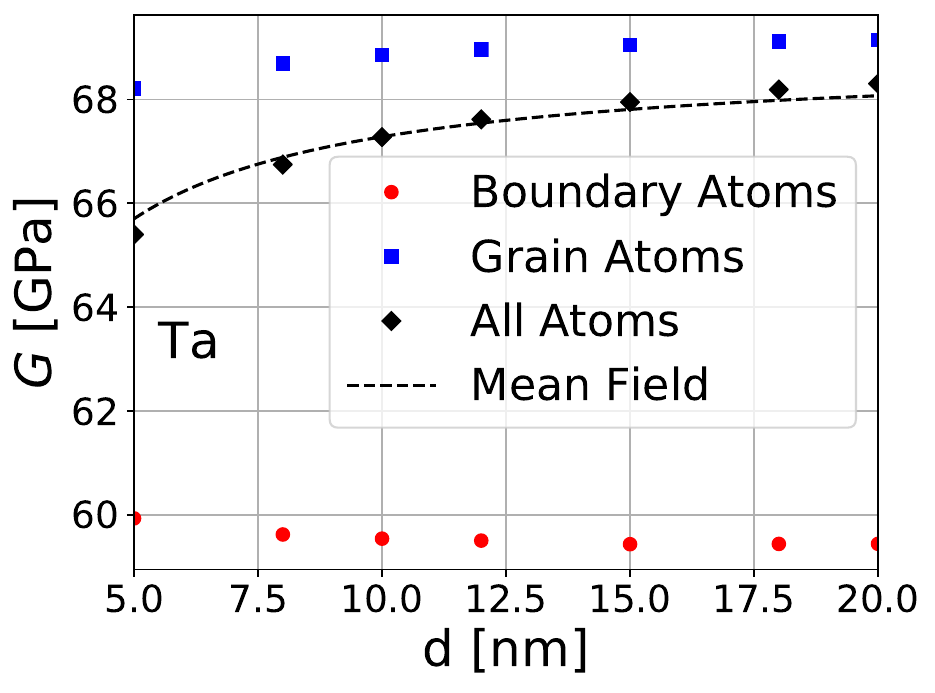}\includegraphics[scale=0.28]{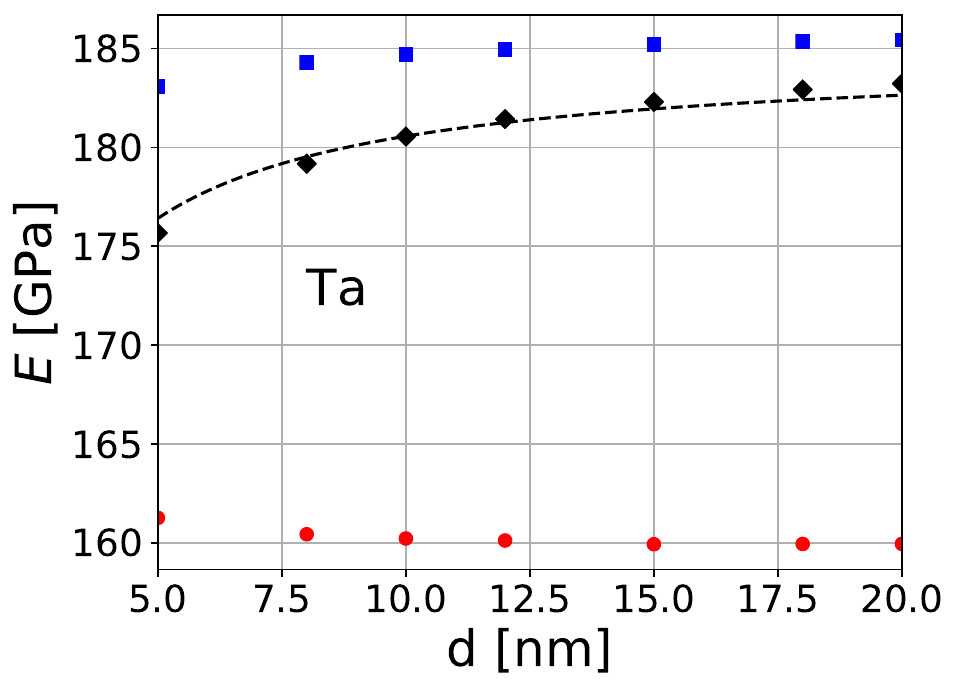}\includegraphics[scale=0.28]{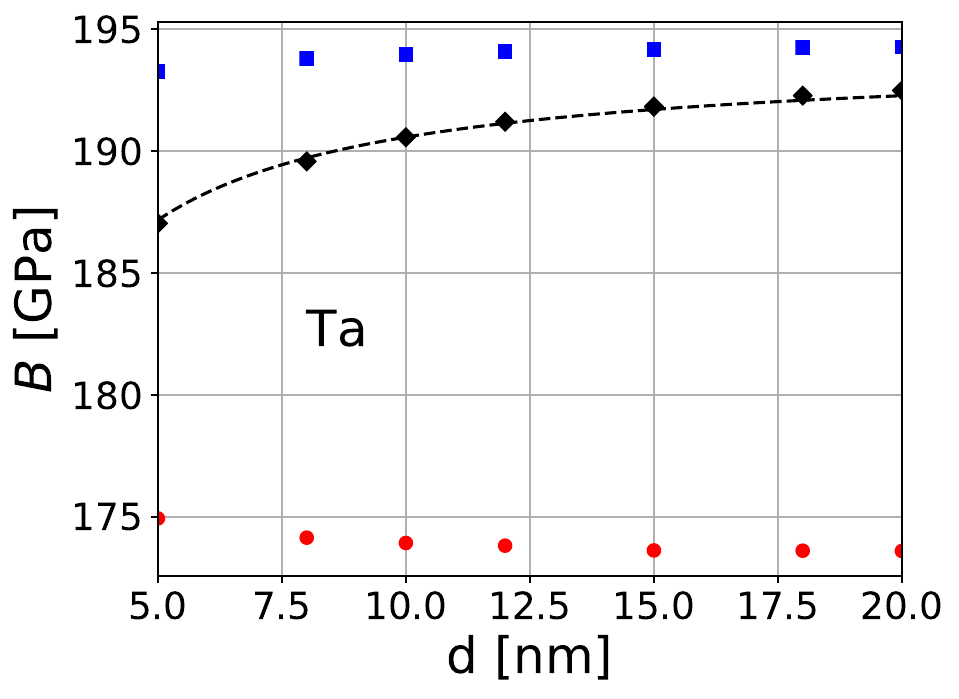}\includegraphics[scale=0.28]{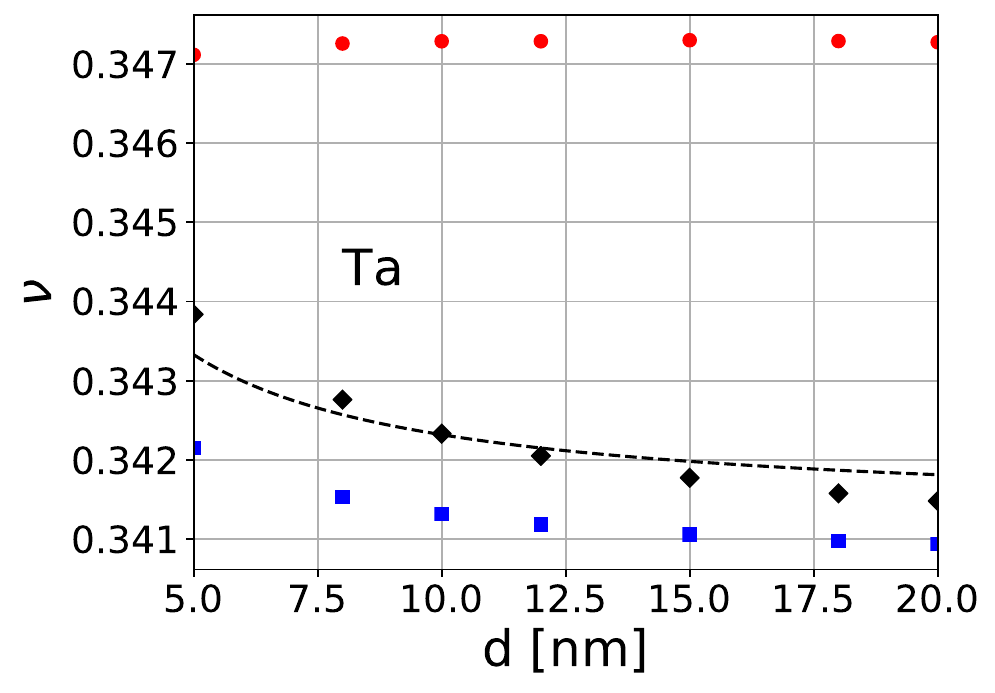} 
\par\end{centering}
\caption{Elastic moduli for grain-boundary (red circles), grain (blue squares) and all
(black diamonds) atoms, as a function of grain size, for nanocrystalline copper (upper figures) and tantalum (lower figures).
From left to right, the shear modulus $G$, Young's modulus $E$, bulk modulus $B$, and Poisson's ratio $v$ are shown.
 The dashed black line represents the mean-field model with   $1/d$ scaling of the grain boundary contribution (Eq. \eqref{eq:Mav_mf}) with $d_0 \approx 1.5 \mathrm{nm}$ for copper and $d_0\approx 1.7 \mathrm{nm}$ for tantalum).
\label{fig:f_d}}
\end{figure*}

Simulations were performed using the state-of-the-art molecular
dynamics code LAMMPS (Large-scale Atomic Molecular Massively Parallel
Simulator) \cite{plimpton1995fast}, to which we added a parallel
implementation of the calculation of the local elasticity tensor for
EAM potentials (equations \eqref{eq:cborn_micro}-\eqref{eq:g_def}).
Results are reported for nanocrystalline copper and tantalum with grain sizes of  $d=5,8,10,12,15,18,20$nm and a corresponding number of atoms 
in the range of $2\times10^{5}-1.5\times10^{7}$.
Copper was simulated using EAM potential by Mishin et al. \cite{mishin2001structural} for an FCC lattice structure with $a=3.615\text{A}$ and tantalum was simulated 
using EAM potential by Ravelo et al. \cite{ravelo2013shock} for a BCC lattice with $a=3.304\text{A}$. Initial configurations were generated by randomly assigning FCC and BCC crystallites on a grain-size super lattice and defining interfaces between grains using Voronoi tessellation. These initial configurations were relaxed using varying box-size NPT molecular dynamics simulations with a target pressure of zero for a duration of $200$ps (with a 2fs time-step), as shown in Figure \ref{fig:p_t_ta}. 
It is evident that as the system reaches the state of equilibrium, the instantaneous pressure oscillates indefinitely with a well defined frequency and amplitude, while the cumulative temporal average pressure decays to zero. This behaviour of the pressure fluctuations is expected from a molecular dynamics simulation in the NVT ensemble (as shown in detail in Ref. \cite{krief2021calculation}, and in the references therein).
The final local pressure distributions in grain and grain boundaries are shown in Figure \ref{fig:p_ta_histograms}.

Final relaxed configurations of several simulation boxes, visualized with Ovito \cite{stukowski2009visualization}
 are shown in Figure \ref{fig:boxes}, where
atoms color are assigned according to the local average centrosymmetry parameter \cite{kelchner1998dislocation,bulatov2006computer},
distinctively separating grain and grain-boundary atoms \cite{hahn2015grain}.

In Figure \ref{fig:snapshots}, planar slices are shown with atoms colored
according to the per-atom shear modulus (Eq. \eqref{eq:G}).

It is evident that the shear modulus varies considerably within grain-boundary regions in contrast with a significantly lower variation in the inter-grain regions. This behavior
is observed for all elastic moduli (defined in equations \eqref{eq:B}-\eqref{eq:poiss}),
and is expected due to the contrast between the amorphous and crystalline atomic arrangements. Histograms for the
per-atom shear modulus of grain and grain-boundary atoms are shown in Figure
\ref{fig:histograms}, for various grain sizes. 
The distributions for intra-grain moduli have a 
 width to average ratio of $\sim5$\%, while the grain-boundary moduli distributions have a ratio close to unity. While this variation is expected, we further examined the dependence of these distributions on grain size. Here, we found that while intra-grain moduli distributions became sharper (lower width over mean) with increasing grain size, such variation was not observed in grain-boundary moduli distributions. 
 Similar behavior is observed for the stress distributions as seen in Figure \ref{fig:p_ta_histograms}.
 We note that even though the average grain-boundary shear modulus is lower than the average grain shear modulus (as expected), it is seen in Figure \ref{fig:histograms} that some grain-boundary local values are larger than local grain values. This is a result of the highly frustrated local amorphous configuration, which results in some grain boundary local values that are higher than those of the ordered phase.

We now discuss the uncertainty in the results. We expect the main variation to be due to the structure of interfaces and triple junctions formed during the Voronoi tessellation process.
In order to quantify this uncertainty, we divided the simulation box into 8 different octants and compared the resulting moduli distributions in the various octants, which contain different grains, interfaces and junctions. We found that the different distributions are very similar, with the average value varying up to 2\%, which can serve as a measure of the uncertainty of the average elasticity values (which are shown below in Fig. \ref{fig:f_d}). This variation does not depend strongly on grain size. This is demonstrated in Fig. \ref{fig:histograms_err}, where we compare the grain and grain boundary shear modulus distributions in 8 different octants and in the entire simulation box, for $d=10$nm nanocrystlline copper.

In Figure \ref{fig:atfrac}, the atomic fraction of grain and grain-boundary atoms are plotted as a function of grain size. It is evident that the grain-boundary atoms fraction
has a strong decreasing dependence on grain size, as expected \cite{schiotz1999atomic,hahn2015grain}. In fact, the number of grain-boundary atoms must scale as the inverse of grain size; that is, if we denote by $x_{g}$ and $x_{gb}$, the grain and grain-boundary atomic
fractions, respectively, then:

\begin{equation}
x_{gb}(d) = \frac{d_0}{d},  \ \ x_{g}(d)=1-\frac{d_0}{d}, \label{eq:n_d_inv}
\end{equation}

where $d_0$ is a constant length scale that depends on the lattice symmetry, and given by $d_0 \approx \frac{n_{gb}}{n_{g}}\Delta_{gb}$ where $\Delta_{gb}$ is the typical grain-boundary width and $n_{gb}, \ n_{g}$ are, respectively, the grain and grain-boundary atom density. These results can be derived from a simplistic spherical grain-boundary model for which $x_{gb}=\frac{4\pi d^{2}\Delta_{gb} n_{gb}}{\frac{4}{3}\pi d^{3} n_{g} }$.
From the data presented in Figure \ref{fig:atfrac}, we find that for copper $d_0 \approx 1.5 \mathrm{nm}$ and for tantalum $d_0\approx 1.7 \mathrm{nm}$.
The scaling law  \eqref{eq:n_d_inv} is also plotted in Figure \ref{fig:atfrac}, which shows that it perfectly matches the simulations' values. Finally, we also note that the
decrease of grain-boundary atomic fraction with grain size $d$, can also be seen
visually in Figures \ref{fig:boxes}-\ref{fig:snapshots}. 

Figure \ref{fig:f_d} presents the grain size dependence
of various elastic moduli, namely,  the shear modulus $G$, Young's modulus $E$ and bulk
modulus $B$, as well as the Poisson's ratio $v$. These moduli are plotted separately for grain and grain-boundary atoms and for the entire system (that is, all atoms in the simulations), resulting in the respective total bulk moduli.

It is evident that
the total elastic moduli increases with grain size,
in accordance with the results of Refs. \cite{latapie2003effect,pan2008tensile,valat2017grain, kowalczyk2020elastic}.
In order to analyze this behavior, we write the total elastic moduli
as a sum of grain and grain-boundary moduli:
\begin{equation}
\mathcal{M}_{\mathrm{total}}\left(d\right)=x_{gb}\left(d\right)\mathcal{M}_{gb}\left(d\right)+x_{g}\left(d\right)\mathcal{M}_{g}\left(d\right), \label{eq:Mav}
\end{equation}
where $\mathcal{M}$ represents one of the considered moduli $G, \ E, \ B, \ v$ and $\mathcal{M}_{\mathrm{total}}, \ \mathcal{M}_{g}$ and $\mathcal{M}_{gb}$
are, respectively, the total, grain and grain-boundary elastic
moduli. The results in Figure \ref{fig:f_d}
indicate that the grains have elastic moduli $\mathcal{M}_{g}$ that are
about 10\% larger than the grain-boundary values $\mathcal{M}_{gb}$.
This is to be expected since it is well-known that the grain boundary
has an amorphous structure \cite{gleiter2000nanostructured,wolf2005deformation,keblinski1999structure},
a result that is consistent with our simulations and with Figure \ref{fig:histograms}
which shows wide elastic modulus distributions for grain-boundary
atoms. Moreover, it is evident that $\mathcal{M}_{g}\left(d\right)$ and $\mathcal{M}_{gb}\left(d\right)$
both have a relatively weak dependence on the grain size $d$. Therefore, we define a "mean-field" model, assuming
 grain-size independent grain and grain-boundary elastic moduli $\mathcal{M}_{g}(d)\approx \overline{\mathcal{M}_{g}}$, $\mathcal{M}_{gb}(d)\approx \overline{\mathcal{M}_{gb}}$, where $ \overline{\mathcal{M}}$ represents arbitrary average of the elastic modulus over grain size $d$. We take the arithmetic average over the grain sizes considered. Using the $1/d$ scaling of the grain-boundary atoms fraction (Eq. \eqref{eq:n_d_inv}), we obtain a simple approximate mean-field model for the exact elastic moduli in Eq. \eqref{eq:Mav}, as a function of grain size:

\begin{equation}
\mathcal{M}^{\mathrm{{Mean-Field}}}_{\mathrm{total}}\left(d\right)= \frac{d_0}{d}\overline{\mathcal{M}_{gb}}+\left(1-\frac{d_0}{d}\right)\overline{\mathcal{M}_{g}},\label{eq:Mav_mf}
\end{equation}

The results of this simple mean field model are also shown in Figure \ref{fig:f_d}, offering a very good qualitative and even quantitative agreement with the exact results obtained from the simulations. Therefore, we deduce that the grain size dependence of the total elastic
moduli $\mathcal{M}_{tot}\left(d\right)$, is mainly due to the strong grain-size dependence 
of the grain and grain-boundary atoms fraction (via. Eq. \eqref{eq:n_d_inv}), while the effect of the grain size dependence of the grain and
grain-boundary elastic moduli is only a second order effect.

\section{Summary}
Local elastic moduli of nanocrystalline copper and tantalum were calculated
using molecular dynamics simulations. The resulting moduli distributions
within the polycrystals were calculated and analyzed as
a function of grain size in the range of $5$-$20$ nm. 
The shear modulus distribution in the amorphous grain boundary
is extended and grain-size independent, while the shear modulus distribution
for atoms within the grains is significantly narrower 
and becomes sharper with increasing grain size. 
Even though local properties do not depend on grain size, global average elastic moduli do show grain size dependence in accordance with previous observations \cite{yoshimoto2004mechanical, papakonstantopoulos2005local, tsamados2009local, mizuno2013measuring}.

This effect is shown explicitly to be denominated by the variation in relative numbers of crystalline versus grain-boundary atoms.
We show for the simulated system that the average global 
elastic moduli can be approximated by a weighted average of the larger
grains elastic moduli and the lower grain-boundary elastic moduli,
an approximation which was assumed in several previous works \cite{kim1999effects, shen1995elastic,latapie2003effect,lian2013emergence, kowalczyk2020elastic}. 

Moreover, by employing the simple inverse relation between the grain-boundary atoms fraction and grain size $d$, we suggest a simple but accurate mean-field formula for the grain size dependence of various bulk atomic moduli (shear modulus, Young's modulus, bulk modulus and Poisson's ratio).

\bibliographystyle{unsrt}
\bibliography{datab}

\pagebreak
\end{document}